\documentclass[aps,prl,reprint,groupedaddress,showpacs]{revtex4-1}
\usepackage{graphicx}

\begin{document}


\title{Localization and circulating currents in curved graphene devices}


\author{ G. M. M. Wakker, Rakesh P. Tiwari and M. Blaauboer}
\affiliation{Delft University of Technology, Kavli Institute of Nanoscience,
Department of Quantum Nanoscience, Lorentzweg 1, 2628 CJ Delft, The Netherlands}


\date{\today}

\begin{abstract}
We calculate the energy spectrum and eigenstates of a graphene sheet which contains a circular deformation. Using time-independent perturbation theory with the ratio of the height and width of the deformation as the small parameter, we find that due to the curvature the wavefunctions for the various states acquire unique angular asymmetry. We demonstrate that the pseudo-magnetic fields induced by the curvature result in circulating probability currents. These circulating currents in turn produce local \textit{real} magnetic fields $\sim$ 100 $\mu$T which can be measured using current technology.

\end{abstract}


\pacs{}


\maketitle

Graphene, a monolayer of carbon atoms in a honeycomb lattice, has attracted a lot of interest in the last decade \cite{g1,g2,g3}. Along with many interesting electronic transport properties \cite{g4} graphene has some intriguing structural and mechanical properties \cite{g5}. A suspended graphene sheet can be deflected by applying a gate voltage \cite{fogler}. Early on, curvature induced by topological defects was studied in the context of carbon nanotubes and fullerenes \cite{ntf,ntf2,gonzalez1,gonzalez2}. Graphene sheets deposited on a substrate naturally show corrugations \cite{meyer,dgs} which can be modeled as defects \cite{Gilbertini}. Deformations in a graphene sheet can provide a mechanism for spin relaxation which is stronger than relaxation due to intrinsic spin orbit interactions \cite{spin}. These deformations can also lead to significant effective pseudo-magnetic fields \cite{nanobub}. Two recent works have investigated properties of pseudo-magnetic fields in graphene sheets: In Ref. \cite{dejuan2} de Juan \emph{et al.} investigate Aharonov-Bohm interferences in the local density of states due to a fictitious strain-induced magnetic field, and in Ref. \cite{abed} Abedpour \emph{et al.} calculate the pseudo-magnetic field induced by shear stress in circular graphene rings. 

The purpose of this Letter is to investigate the possibility of exploiting curvature, induced by elastic deformations, in a graphene sheet for generating states with circulating currents. We focus on systems where the curvature in the graphene sample can be controlled externally. A possible way of realizing such systems is to suspend graphene samples on a substrate where an annulus has been etched away. Modeling the curvature as a 
Gaussian-shaped bump and using perturbation theory, we show that it is possible to generate localized circulating carriers in the ground state for such systems.


According to the standard theory of elasticity \cite{ando,manes,dejuan,Gilbertini} the effect of curvature can be captured by including a scalar ($V_1$) and vector ($V_2=A^x_{ps}-iA^y_{ps}$) potential in the Hamiltonian \cite{Gilbertini} with
$V_1=g_1(u_{xx}+u_{yy})$ and $V_2=g_2(u_{xx}-u_{yy}+2iu_{xy})$. Here $u_{ij}$ (with $i,j \in \{x,y\}$) is the usual deformation tensor,
\begin{equation}
u_{ij}=\frac{1}{2}\left(\frac{\partial \bar{u}_i}{\partial x_{j}}+\frac{\partial \bar{u}_j}{\partial x_{i}}+\sum_{k\in\{x,y,z\}}\frac{\partial \bar{u}_k}{\partial x_{i}}\frac{\partial \bar{u}_k}{\partial x_{j}}\right),
\end{equation} 
with $\bar{u}_i=\bar{u}_i({\bf r})$, $i\in\{x,y,z\}$, the Cartesian components of the average displacements. The coupling constants $g_1$ and $g_2$ are known from transport measurements \cite{Gilbertini}.  
\begin{figure}
\includegraphics[scale=.285]{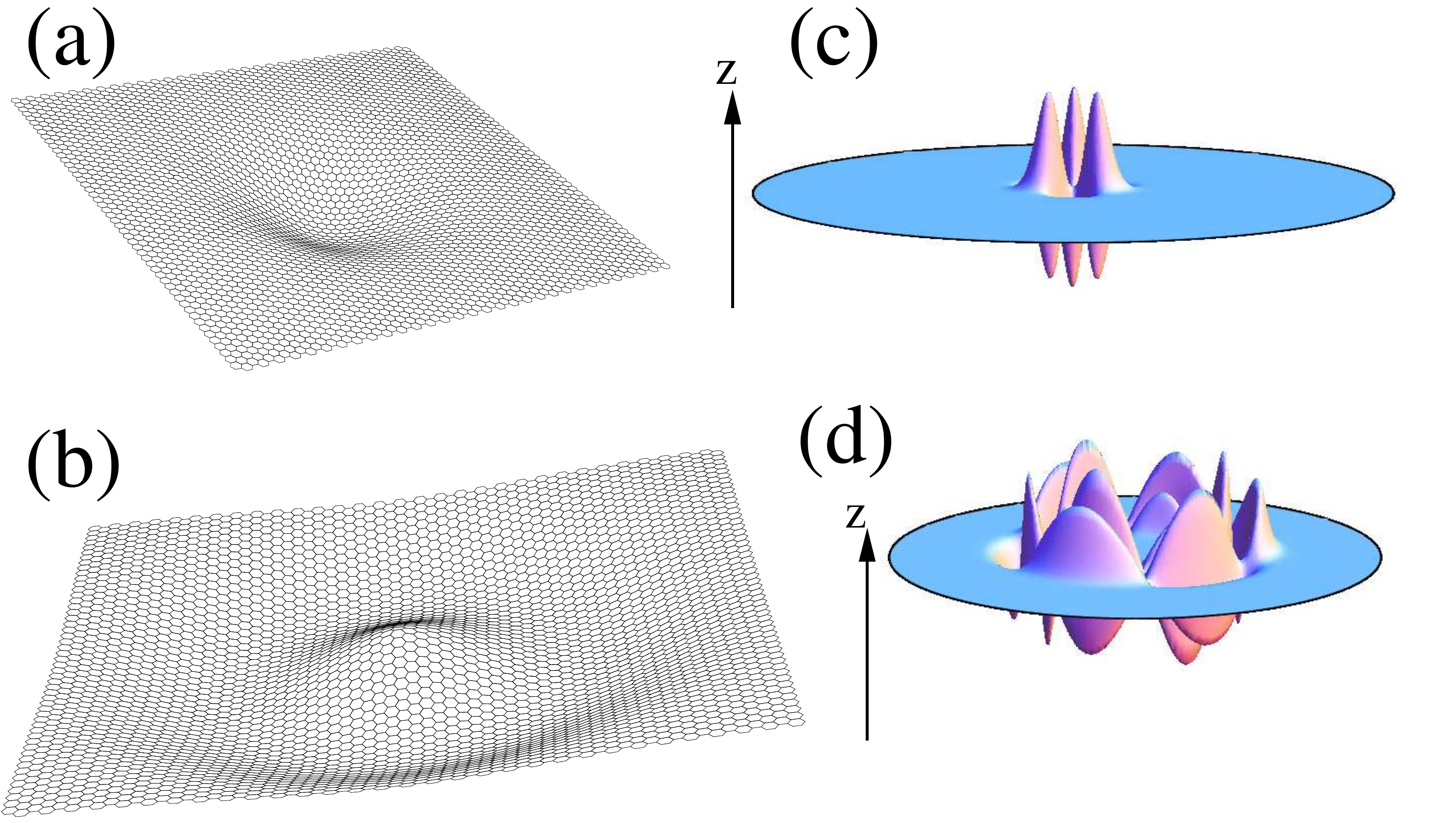}
\caption{(Color online) (a) Schematic of type (a) deformation. (b) Schematic of type (b) deformation. (c) Pseudo-magnetic field produced by type (a) deformation. (d) Pseudo-magnetic field produced by type (b) deformation. See the text for further details. }
\label{fig:Gvsalpha}
\end{figure}
We assume that the length scales associated with the deformation are much larger than $a_{cc}$, where $a_{cc}\sim 1.4$ \AA represents the lattice constant of graphene, so that inter valley scattering can be neglected. The single-valley Dirac Hamiltonian for the system in the absence of external magnetic fields is given by: 
\begin{equation}
\mathcal{H}=v_F\left(\begin{array}{cc}
-eV_1 & p_x-ip_y+ eV_2 \\
p_x+ip_y+eV_2^{\ast} & -eV_1
\end{array}\right).
\end{equation}
Here $v_F$ is the Fermi velocity and {\bf p}=$-i\hbar{\bf \nabla}$ represents the momentum operator.
Now we focus on the shape of the deformation. In particular we consider two deformations: (a) a Gaussian centered at the origin, i.e $u_z({\bf r})=\mathcal{A}e^{-\frac{x^2+y^2}{b^2}}$ and (b) a ring-shaped Gaussian deformation modeled as $u_z({\bf r})=\mathcal{A}e^{-\frac{(x-cx/\sqrt{x^2+y^2})^2+(y-cy/\sqrt{x^2+y^2})^2}{b^2}}=\mathcal{A}e^{-\frac{(\sqrt{x^2+y^2}-c)^2}{b^2}}$. These deformations are schematically depicted in Fig. 1(a) and 1(b). We consider deformations with small height and large width and apply perturbation theory, using $\alpha\equiv\mathcal{A}/b$ as the small parameter. In polar coordinates the potentials $V_1$ and $V_2$ are given by
$V_1=2g_1\frac{\mathcal{A}^2}{b^2}f(r)$ and $V_2=2g_2\frac{\mathcal{A}^2}{b^2}e^{i2\theta}f(r)$, where 
\begin{equation}
f(r)=\left\{\begin{array}{c}
\frac{r^2}{b^2}e^{-2r^2/b^2} \textrm{ for type (a) deformations} \\
\frac{(r-c)^2}{b^2}e^{-2(r-c)^2/b^2} \textrm{ for type (b) deformations. }
\end{array}\right.
\end{equation}
The Hamiltonian [Eq. (2)] can be written in polar coordinates as $\mathcal{H}=\mathcal{H}_0+\mathcal{H}_1$ with
\begin{equation}
\mathcal{H}_0=-i \hbar v_F \left(\begin{array}{cc}
0 & e^{-i\theta}\left(\partial_r - \frac{i}{r}\partial_\theta \right) \\
e^{i\theta}\left(\partial_r + \frac{i}{r}\partial_\theta \right) & 0
\end{array}
\right),
\end{equation}
and
\begin{equation}
\mathcal{H}_1=2e\alpha^2f(r)\left(\begin{array}{cc}
-g_1 & g_2e^{2i\theta} \\
g_2e^{-2i\theta} & -g_1
\end{array}
\right).
\end{equation}

\begin{figure}
\includegraphics[scale=.43]{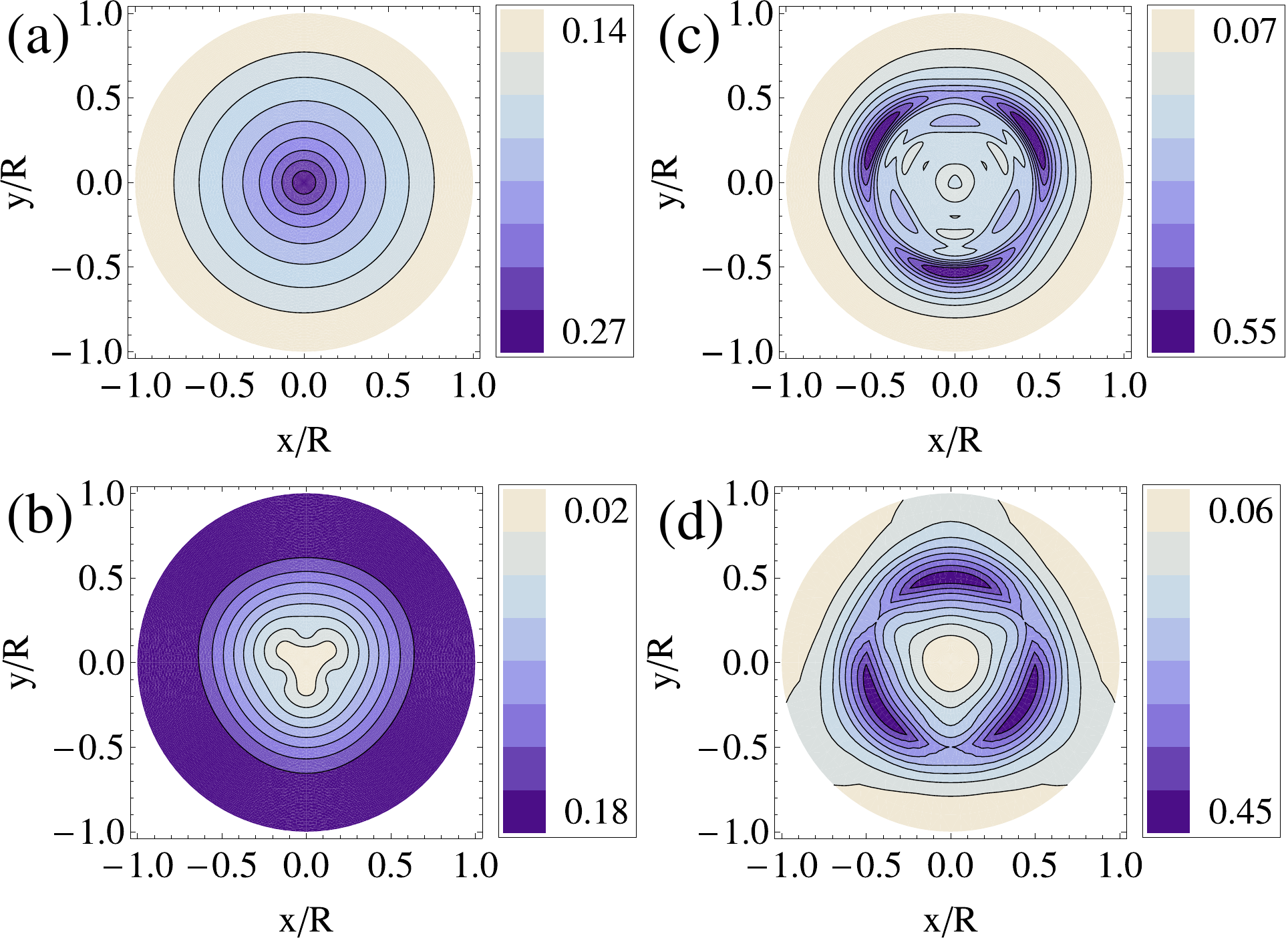}
\caption{(Color online) Contour plot of $ \Psi_{k_{m,l}}^{\dagger}\Psi_{k_{m,l}}$ (a) for the lowest electronic state of type (a) deformation, (b) for the first excited state of type (a) deformation, (c) for the lowest electronic state of type (b) deformation and (d) for the first excited state of type (b) deformation. Parameters used are $\alpha=0.04$, $\beta=0.1$, $\gamma=0.4$, $eg_1=3 $\, eV, $eg_2=2.276 $\, eV \cite{Gilbertini}. See the text for further details.}
\label{fig:2}
\end{figure}
Here $\mathcal{H}_0$ describes the unperturbed system, which is a flat circular graphene sheet of radius $R$ and $\mathcal{H}_1$ is the perturbation. The normalized eigenstates of the unperturbed Hamiltonian $\mathcal{H}_0$ are given by \cite{peres}:
\begin{equation} 
\Psi_{k,m}(r,\theta)=\sqrt{\frac{1}{2}}\left(\begin{array}{c}
J_{m}(r k)e^{i m \theta} \\
is J_{m+1}(r k)e^{i(m+1)\theta}
\end{array}\right),
\end{equation}
where $J_{m}(rk)$ is the Bessel function of integer order $m$ with energies $\epsilon_{m,k}=s\hbar v_F k$. $s=\pm 1$ represents the conduction and the valence band. It should be noted that the states described by Eq. (6) are also eigenfunctions of the $z$-component of the total angular momentum $\hat{L}_z$:
\begin{equation}
\hat{L}_z \Psi_{k,m}=(-i\hbar\partial_{\phi} + \frac{\hbar}{2}\hat{\sigma}_{z})\Psi_{k,m}=(m+\frac{1}{2})\Psi_{k,m}. 
\end{equation}
Now we impose infinite mass boundary conditions \cite{berry} which are valid in the regime $R\gg a_{cc}$. This can be done individually for each $m$ and results in the eigenvalue equation:
\begin{equation}
J_{m}(k_{m,l} R)=J_{m+1}(k_{m,l} R),
\end{equation} 
where the integer $l$ represents the ascending order of the values $\mid k_{m,l}\mid$ that satisfy the above equation.
The eigenvalues for $m$ and $-m-1$, which correspond to equal and opposite total angular momenta, are not degenerate. The spectrum resulting from Eq. (8) is discrete due to the finiteness of the graphene sheet, and exhibits a band gap at the Dirac point ($k_{m,l}=0$). The solution of Eq. (8), closest to the Dirac point is given by $k_{-1,-1}= -\frac{1.4347}{R}$ and $k_{0,1}=\frac{1.4347}{R}$. The magnitude of the band gap is thus $2\hbar v _F \frac{1.4347}{R}\approx 30$ meV for $R=50$ nm \cite{bandgap}. It is useful to write the perturbation Hamiltonian $\mathcal{H}_1$ as
\begin{equation}
\mathcal{H}_1=\frac{2e\alpha^2}{\beta^2}f(x)\left(\begin{array}{cc}
-g_1 & g_2e^{2i\theta} \\
g_2e^{-2i\theta} & -g_1
\end{array}
\right),
\end{equation}
where $\beta\equiv b/R$ and $x=r/R$. The form of the deformation is now given as
\begin{equation}
\beta^2f(x)=\left\{\begin{array}{c}
x^2e^{-2\frac{x^2}{\beta^2}} \textrm{ for type (a) deformations} \\
(x-\gamma)^2e^{-2\frac{(x-\gamma)^2}{\beta^2}} \textrm{ for type (b) deformations. }
\end{array}\right.
\end{equation}
Here $\gamma \equiv c/R$ is a measure of the location of the deformation for type (b) deformations. The vector potential in our formalism generates a pseudo-magnetic field ${\bf B}_{ps} = \nabla \times {\bf A}_{ps}$, perpendicular to the graphene sheet. For type (a) deformations this magnetic field is
${\bf B}_{ps}^{(a)}=8g_2\frac{r^3}{b^4}e^{-2\frac{r^2}{b^2}}\sin(3\theta) \hat{z}$ as sketched in Fig. 1(c). Similarly, for type (b) deformations ${\bf B}_{ps}^{(b)}=\frac{4g_2}{b^4r}e^{-2\frac{(r-c)^2}{b^2}}(6c^2r^2+2r^4-2c^3r-6cr^3+b^2c^2-b^2cr)\sin(3\theta) \hat{z}$
as shown in Fig. 1(d). The pseudo-magnetic fields are not \textit{real} magnetic fields but a consequence of the deformation. Although these fields are zero when averaged over the entire circular graphene sheet (as can also be seen from Figs. 1(c) and 1(d)), they can locally be quite large, see the discussion at the end of the Letter.

Now we calculate the corrections due to the curvature to the energy spectrum and eigenfunctions up to second order in $\alpha^2$. The energy  eigenvalues are then given by $\epsilon_{k_{m,l}}=\epsilon^{(0)}_{k_{m,l}}+\epsilon^{(1)}_{k_{m,l}}+\epsilon^{(2)}_{k_{m,l}}$. Similarly, the wavefunctions are given by $\Psi_{k_{m,l}}=\Psi_{k_{m,l}}^{(0)} + \Psi_{k_{m,l}}^{(1)}$. The unperturbed eigenvalues $\epsilon^{(0)}_{k_{m,l}}$ and wavefunctions $\Psi_{k_{m,l}}^{(0)}$ are given by $\epsilon^{(0)}_{k_{m,l}}=s\hbar v_F k_{m,l}$ and 
\begin{equation}
\Psi_{k_{m,l}}^{(0)}=A_{ml}\left(\begin{array}{c} J_{m}(r k_{m,l})e^{i m \theta} \\ is J_{m+1}(r k_{m,l})e^{i(m+1)\theta}\end{array}\right), 
\end{equation}
where $A_{ml}$ is the normalization factor given by
\begin{equation}
A_{ml}=2\int_{0}^{1}\left[J_{m}^2(k_{m,l}x)+J_{m+1}^2(k_{m,l}x)\right]xdx.
\end{equation}
The corrections $\epsilon^{(1)}_{k_{m,l}}$, $\epsilon^{(2)}_{k_{m,l}}$ and $\Psi_{k_{m,l}}^{(1)}$ are calculated by standard time-independent non-degenerate perturbation theory \cite{sakurai} and we obtain:
\begin{eqnarray}
\epsilon^{(1)}_{k_{m,l}}&=& -\frac{4e\alpha^2 g_1}{\beta^2}\frac{\int_{0}^{1}f(x)\left[J_{m}^2(k_{m,l}x)+J_{m+1}^2(k_{m,l}x)\right]xdx}{A_{ml}^2} \nonumber \\
\epsilon^{(2)}_{k_{m,l}}&=&(2e\alpha^2)^2\sum_{k_{m^{\prime}l^{\prime}}\neq k_{m,l}}\frac{\mid\mathcal{V}_{k_{m^{\prime}l^{\prime}},k_{m,l}}\mid^2}{\epsilon^{0}_{k_{m,l}}-\epsilon^{0}_{k_{m^{\prime}l^{\prime}}}}
\nonumber \\
\Psi_{k_{m,l}}^{(1)}&=&2e\alpha^2\sum_{k_{m^{\prime},l^{\prime}}\neq k_{m,l}} \frac{\mathcal{V}_{k_{m^{\prime},l^{\prime}};k_{m,l}}}{\epsilon^{0}_{k_{m,l}}-\epsilon^{0}_{k_{m^{\prime},l^{\prime}}}}\Psi^{(0)}_{k_{m^{\prime},l^{\prime}}}.
\label{eq:corrections}
\end{eqnarray}
Here the matrix element $\mathcal{V}_{k_{m^{\prime},l^{\prime}};k_{m,l}}$ is given by:
\begin{widetext}
\begin{eqnarray}
\mathcal{V}_{k_{m^{\prime},l^{\prime}};k_{m,l}}&=&\frac{1}{A_{m^{\prime}l^{\prime}}A_{ml}\pi \beta^2}\int_{0}^{2\pi}\int_{0}^{1}e^{i(m-m^{\prime})\theta}\left[J_{m}(k_{m,l}x), -is J_{m+1}(k_{m,l}x)e^{-i\theta}\right] \nonumber \\
& & \left[\begin{array}{cc}
-g_1 & g_2e^{2i\theta} \\
g_2e^{-2i\theta} & -g_1
\end{array}
\right]\left[\begin{array}{c} J_{m^{\prime}}(k_{m^{\prime},l^{\prime}}x) \\ is J_{m^{\prime}+1}(k_{m^{\prime},l^{\prime}}x)e^{i\theta}\end{array}\right]x\, dx\, d\theta.
\label{eq:matrixelement}
\end{eqnarray}
\end{widetext}
The angular integral in Eq.~(\ref{eq:matrixelement}) is only non-zero if $m=m^{\prime}$, $m=m^{\prime}+3$ or $m=m^{\prime}-3$, which simplifies the calculations. 

Now we present our numerical results. In principle the sum in the expressions for $\epsilon^{(2)}_{k_{m,l}}$ and $\Psi_{k_{m,l}}^{(1)}$ [Eqns.~(\ref{eq:corrections})] runs over an infinite number of values of $k_{m,l}$, but we restrict the sum to values of $k_{m,l}$ such that $m\in(-6,5)$ and $l\in(-15,15)$. We have verified that the results obtained by using this restricted sum presented for the lowest two electron-like and lowest two hole-like eigenvalues and eigenfunctions are converged up to at least four significant digits. 


In Fig. 2 we plot the probability density $\Psi_{k_{m,l}}^{\dagger}\Psi_{k_{m,l}}$ over the entire graphene sheet for the lowest electronic state (i.e $k_{m,l}=1.4347/R$) and the first excited electronic state (i.e $k_{m,l}=2.62987/R$) for both type (a) and (b) deformations.
A contour plot of $ \Psi_{k_{m,l}}^{\dagger}\Psi_{k_{m,l}}$ is shown in Fig. 2(a) for the lowest electronic state (ground state) for a central Gaussian [type (a)] deformation. We see that there is no angular dependence of the probability density, which is in agreement with previously obtained results \cite{dejuan,Gilbertini}. 
In contrast, Fig. 2(b) shows that $ \Psi_{k_{m,l}}^{\dagger}\Psi_{k_{m,l}}$ for the first excited electronic state of a type (a) deformation is angle-dependent. This angular dependence is in agreement with the pseudo-magnetic field  ${\bf B}_{ps}^{(a)}$ generated by such a deformation.
For a ring-shaped Gaussian, on the other hand, it is evident from Figs. 2(c) and 2(d) that the probability density $\Psi_{k_{m,l}}^{\dagger}\Psi_{k_{m,l}}$ is angle-dependent for both the ground state and the first excited state. The fact that also the probability density of the ground state is angle-dependent makes the ring-shaped deformations more suitable for experimental demonstration. 
It should be noted that the probability density is high around the location of the deformation ($r = 0.4 R$). As we increase $\alpha$, the effect of deformation-induced \textit{localization} of the wavefunctions for both types of deformations increases. 
\begin{figure}
\centering
\includegraphics[scale=.25]{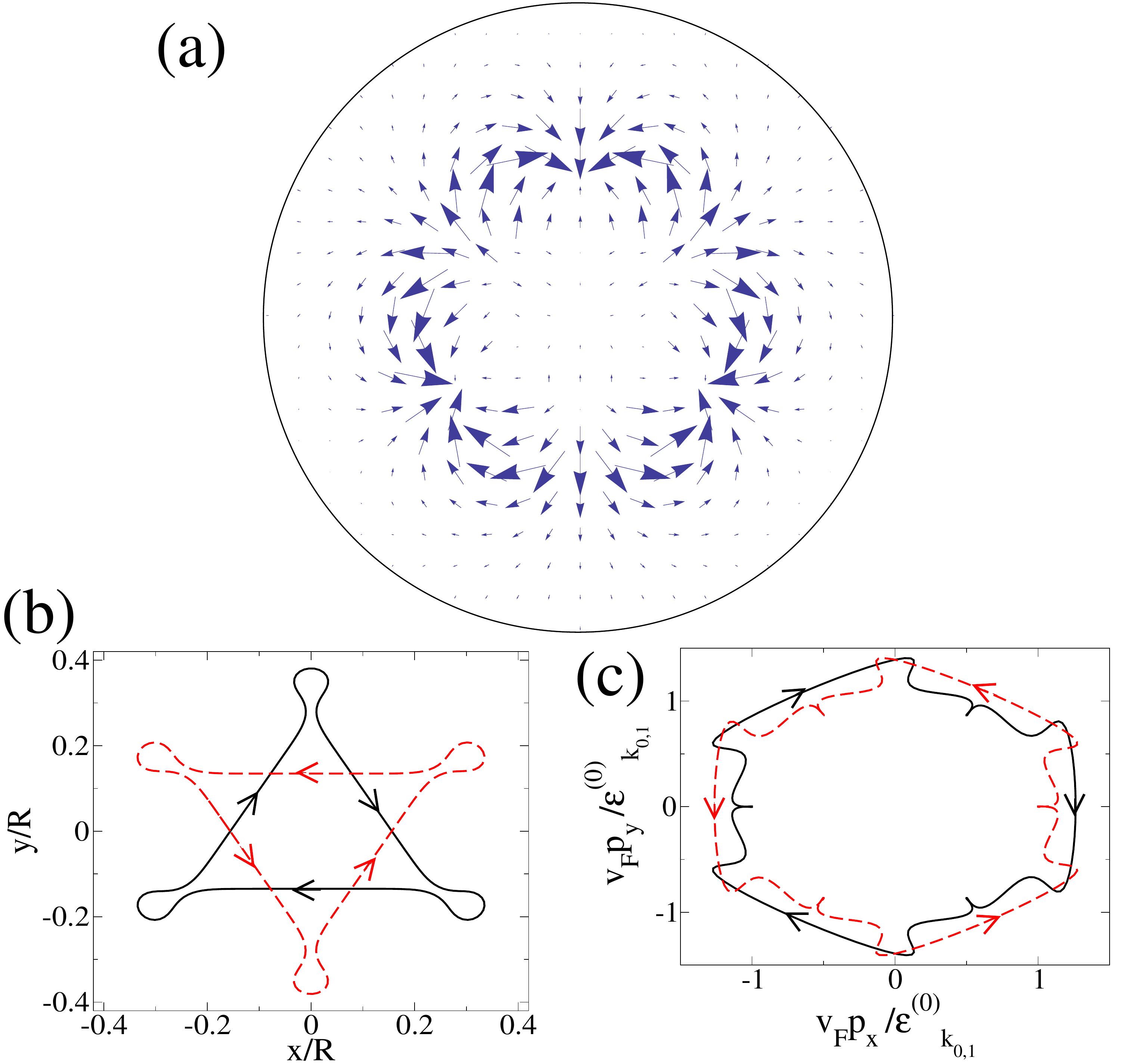}

\caption{(Color online) (a) Vector field plot of ${\bf j}$ over the entire graphene sheet for the lowest electronic state for a ring-shaped Gaussian [type (b)] deformation. (b) Classical trajectory in real space calculated using Eq. (15). Solid (black) and dashed (red) lines represent two different sublattices. (c) Classical trajectory in momentum space calculated using Eq. (15). Again solid (black) and dashed (red) lines represent two different sublattices. See the text for further details. }
\label{fig:4}
\end{figure}
A localized probability density results in circulating currents which in turn generate local magnetic fields. In order to investigate this effect we calculate the probability current density ${\bf j}\equiv v_F \Psi_{k_{m,l}}^{\dagger}{\bf \sigma}\Psi_{k_{m,l}}$. Fig. 3(a) shows a vector field plot of ${\bf j}$ over the entire graphene sheet for the lowest electronic state of a ring-shaped Gaussian [type (b)] deformation.

Localization of the carriers is a repercussion of the topological effect of the pseudo-magnetic field. To elucidate this further we plot in Fig.~\ref{fig:4} the classical trajectories calculated from the corresponding classical Hamiltonian 
\begin{equation}
\mathcal{H}_{class}=\pm v_F\sqrt{(p_x+\frac{A_{ps}^x}{v_F})^2+(p_y+\frac{A_{ps}^y}{v_F})^2}.
\end{equation}
Fig 3(b) shows the position space cyclotron orbits for the ground state of the ring shaped Gaussian deformation \cite{yaroslav}. The two different 
sublattices of graphene have different orbits as is shown by the solid and the dashed lines. Fig.~\ref{fig:4}(c) shows the corresponding momentum space closed orbits. These classical orbits exhibit remarkable similarity to the quantum-mechanical probability density and the probability density current shown in Figs.~\ref{fig:2}(c) and \ref{fig:4}(a). These circulating orbits produce local magnetic fields (which are zero when averaged over the entire graphene sheet). It should be stressed that the existence of these magnetic fields does require the carriers to be valley-polarized, since otherwise the degenerate carriers in the other valley will cancel these local magnetic fields exactly. As has been suggested recently, a special kind of line defect can be used in graphene as a valley filter \cite{gun}. Alternately, weak intervalley scattering has been shown to lift the valley degeneracy \cite{dejuan2}.  
Experimentally one can measure the angular distribution of the probability density using a scanning tunneling microscope (STM). 
To get an estimate for the order of magnitude of the magnetic fields produced we approximate the current density depicted in Fig.~\ref{fig:4}(a) by a circular wire of radius $0.4 R$, where six equal arcs on the circumference carry alternating counter circulating constant currents. For typical carrier densities of $10^{12}$ cm$^{-2}$ the resulting magnetic fields are then of order $100 \mu$T. Magnetic fields of this order can be easily measured by using a single nitrogen-vacancy (NV) impurity in diamond \cite{nv}.

To summarize, we have calculated the energy spectrum and wavefunctions for two circular Gaussian shaped deformations in a graphene sheet using perturbation theory. Due to the curvature in the graphene sheet the carriers become localized around the deformation and follow circulating orbits. These circulating carriers produce local magnetic fields which are zero when averaged over the entire sample. Interesting angular dependence of the probability density is predicted for the ground state of a ring shaped Gaussian deformation. We suggest different ways to experimentally verify our predictions with current technologies, which will provide further insight.   


\begin{acknowledgments}
We thank Ya. M. Blanter and F. Konschelle for stimulating discussions. This research was supported by the Dutch Science Foundation NWO/FOM.
\end{acknowledgments}

\end{document}